\documentclass[prd,reprint,nofootinbib,notitlepage,aps,tightenlines,
preprintnumbers,amsmath,amssymb,amsfonts,showpacs,superscriptaddress]{revtex4-1}

\usepackage[hyperindex,breaklinks]{hyperref}
\usepackage{graphicx,epstopdf}

\usepackage{natbib,bm}
\usepackage[usenames,dvipsnames]{color}
\usepackage{slashed}    
\usepackage{hyperref}
\usepackage{breakurl}
\usepackage{multirow}
\usepackage{bbold}

\def\be{\begin{equation}}
\def\ee{\end{equation}}
\def\ba{\begin{eqnarray}}
\def\ea{\end{eqnarray}}
\def\ge{\mathrel{\raise.3ex\hbox{$>$\kern-.75em\lower1ex\hbox{$\sim$}}}}
\def\la{\mathrel{\raise.3ex\hbox{$<$\kern-.75em\lower1ex\hbox{$\sim$}}}}

\def\simgt{\mathrel{\raise.3ex\hbox{$>$\kern-.75em\lower1ex\hbox{$\sim$}}}}
\def\simlt{\mathrel{\raise.3ex\hbox{$<$\kern-.75em\lower1ex\hbox{$\sim$}}}}

\newcommand{\nc}{\newcommand}

\nc{\gone}{\bar g_{\pi NN}^{(1)}}
\nc{\gzero}{\bar g_{\pi NN}^{(0)}}
\nc{\al}{\alpha}
\nc{\ga}{\gamma}
\nc{\de}{\delta}
\nc{\ep}{\epsilon}
\nc{\ze}{\zeta}
\nc{\et}{\eta}
\nc{\ka}{\kappa}
\nc{\rh}{\rho}
\nc{\si}{\sigma}
\nc{\ta}{\tau}
\nc{\up}{\upsilon}
\nc{\ph}{\phi}
\nc{\ch}{\chi}
\nc{\ps}{\psi}
\nc{\om}{\omega}
\nc{\Ga}{\Gamma}
\nc{\De}{\Delta}
\nc{\La}{\Lambda}
\nc{\Si}{\Sigma}
\nc{\Up}{\Upsilon}
\nc{\Ph}{\Phi}
\nc{\Ps}{\Psi}
\nc{\Om}{\Omega}
\nc{\ptl}{\partial}
\nc{\del}{\nabla}
\nc{\ov}{\overline}
\nc{\newcaption}[1]{\centerline{\parbox{15cm}{\caption{#1}}}}
\nc{\us}{U(1)$_S$}

\def\beq{\begin{equation}}
\def\eeq{\end{equation}}
\def\bmat{\begin{displaymath}}
\def\emat{\end{displaymath}}
\def\bear{\begin{eqnarray}}
\def\eear{\end{eqnarray}}
\def\ba{\begin{eqnarray}}
\def\ea{\end{eqnarray}}
\def\bery{\begin{array}}
\def\ery{\end{array}}
\def\bit{\begin{itemize}}
\def\eit{\end{itemize}}
\def\ben{\begin{enumerate}}
\def\een{\end{enumerate}}
\def\btab{\begin{tabular}}
\def\etab{\end{tabular}}
\def\btbl{\begin{table}}
\def\etbl{\end{table}}
\def\bfig{\begin{figure}[htb]}
\def\efig{\end{figure}}
\def\bpic{\begin{picture}}
\def\epic{\end{picture}}

\def\ga{\mathrel{\raise.3ex\hbox{$>$\kern-.75em\lower1ex\hbox{$\sim$}}}}
\def\la{\mathrel{\raise.3ex\hbox{$<$\kern-.75em\lower1ex\hbox{$\sim$}}}}
\def\gappeq{\mathrel{\rlap {\raise.5ex\hbox{$>$}}
{\lower.5ex\hbox{$\sim$}}}}
\def\lappeq{\mathrel{\rlap{\raise.5ex\hbox{$<$}}
{\lower.5ex\hbox{$\sim$}}}}

\def\gyr{{\rm \, G\kern-0.125em yr}}
\def\mev{{\rm \, Me\kern-0.125em V}}
\def\gev{{\rm \, Ge\kern-0.125em V}}
\def\tev{{\rm \, Te\kern-0.125em V}}

\begin{document}

\title{Light new physics in coherent neutrino-nucleus scattering experiments}

\author{Patrick deNiverville}
\affiliation{Department of Physics and Astronomy, University of Victoria, 
Victoria, BC V8P 5C2, Canada}

\author{Maxim Pospelov}
\affiliation{Department of Physics and Astronomy, University of Victoria, 
Victoria, BC V8P 5C2, Canada}
\affiliation{Perimeter Institute for Theoretical Physics, Waterloo, ON N2J 2W9, 
Canada}

\author{Adam Ritz}
\affiliation{Department of Physics and Astronomy, University of Victoria, 
Victoria, BC V8P 5C2, Canada}

\date{May 2015}

\begin{abstract}
\noindent 
Experiments aiming to detect coherent neutrino-nucleus scattering present opportunities to probe new light weakly-coupled states, such as sub-GeV mass dark matter, in several extensions of the Standard Model. These states can be produced along with neutrinos in the collisions of protons with the target, and 
their production rate can be enhanced if there exists a light mediator produced on-shell. We analyze the sensitivity reach of 
several proposed experiments to light dark matter interacting with the Standard Model via a light vector mediator coupled to the 
electromagnetic  current. We also determine the corresponding sensitivity to massless singlet neutrino-type states with interactions 
mediated by the baryon number current. In both cases we observe that proposed coherent neutrino-nucleus scattering experiments, such as {\tt COHERENT} at the SNS and {\tt CENNS} at Fermilab, will have sensitivity well beyond the existing limits.

\end{abstract}
\maketitle

\section{Introduction}

The cosmic neutrino background constitutes an example of relic dark matter (DM). Although relativistic at decoupling, and playing a sub-dominant role in structure formation, neutrinos are otherwise characteristic of a cosmic relic that can undergo elastic scattering with nuclei. While the small mass and low temperature of the cosmic neutrino background makes this scattering channel a very challenging target for detection, the signature of coherent neutrino-nucleus scattering, e.g. by solar, reactor or accelerator neutrinos, served as the initial template for the 
signatures used in underground dark matter detectors \cite{ds,gw}. An improvement by roughly three orders of magnitude in the sensitivity  
of underground DM detectors will bring these experiments to the detection threshold of the solar neutrino flux, complicating
further progress in direct dark matter searches.  

Coherent elastic neutrino-nucleus scattering (CE$\nu$NS) is, of course, of considerable interest in its own right. An accurate determination of the cross section may allow an independent measurement of the low-energy value of the weak mixing angle $\theta_W$. Moreover, deviations from the expected rate may signal physics beyond the Standard Model (SM) in the neutrino sector, often parametrized as NSI - non-standard neutrino interactions (see {\em e.g.} \cite{Scholberg:2005qs}). 
Detecting CE$\nu$NS would also showcase the progress made in the development of detector technologies used for dark matter detection, and in searches for neutrinoless double-beta decay.  The experimental activity directed towards CE$\nu$NS has intensified recently. Besides the possibility of using reactor neutrinos \cite{Moroni:2014wia}, a stopped-pion source of neutrinos can be used to detect CE$\nu$NS. At the moment, the BNB at Fermilab and the SNS at Oak Ridge, 
represent realistic opportunities, with experimental proposals under active consideration \cite{Akimov:2013yow,Brice:2013fwa,Collar:2014lya}. 

With the close analogy between CE$\nu$NS and dark matter scattering in mind, it should not be a surprise that proposals to measure coherent neutrino scattering in high luminosity fixed target experiments are also well suited to searching for new light states other than the SM neutrinos. 
In this paper, we will show that modern fixed target proposals designed to observe and measure CE$\nu$NS are also very sensitive to generic models of light dark matter in the 1--100 MeV mass range. It has been appreciated that fixed target experiments provide a complementary approach to direct dark matter detection, with superior sensitivity in the sub-GeV mass range. The use of high intensity proton-beam fixed target neutrino oscillation experiments \cite{Batell:2009di,deNiverville:2011it,deNiverville:2012ij,Kahn:2014sra} and also electron-beam fixed target experiments \cite{Izaguirre:2013uxa,Diamond:2013oda,Izaguirre:2014dua,Batell:2014mga} has recently been highlighted as a means to probe the light dark matter parameter space, 
and a dedicated beam-dump run was recently carried out at the MiniBooNE experiment \cite{Dharmapalan:2012xp}. (See also \cite{Hewett:2012ns,Kronfeld:2013uoa,Essig:2013lka,pospelov2008,Batell:2009yf,Essig:2009nc,Reece:2009un,Bjorken:2009mm,Freytsis:2009bh,Batell:2009jf,Freytsis:2009ct,Essig:2010xa,Essig:2010gu,McDonald:2010fe,Williams:2011qb,Abrahamyan:2011gv,Archilli:2011zc,Lees:2012ra,Davoudiasl:2012ag,Kahn:2012br,Andreas:2012mt,Essig:2013vha,Davoudiasl:2013jma,Morrissey:2014yma,Babusci:2014sta} for studies of related hidden sectors.) In order not to over-produce light dark matter in the early universe, consistent models generically require a relatively light force carrier mediating the interaction between the SM and dark sector, and providing an efficient annihilation channel \cite{Boehm:2003hm}. The light mass scale of the mediator, in turn, increases the production rate of light DM in fixed target collisions. 

Our primary goal in this paper will be to determine the generic sensitivity of facilities designed to measure coherent neutrino scattering to light dark matter. 
To this end, we will study a simple benchmark light dark mater model, with a light vector mediator kinetically mixed with the photon. This class of models is well motivated on effective field theory grounds since 
a kinetically mixed vector is one of the few renormalizable portal couplings to a neutral hidden sector. These models also exemplify the constrained scenarios, which are viable from a phenomenological and cosmological perspective~\cite{Pospelov:2007mp,deNiverville:2012ij,Izaguirre:2015yja}, and have an interaction strength exceeding the weak interactions. We analyze the sensitivity of CE$\nu$NS experiments to production of the vector mediator in the target through both pion decay in flight and charged pion capture, with subsequent on or off-shell decay to dark matter which then scatters coherently (or incoherently) off nuclei in the target. We show that experiments such as {\tt COHERENT} proposed at the SNS, and {\tt CENNS} proposed at Fermilab, can probe significant regions of parameter space that are inaccessible in other searches.

Besides (light) dark matter, CE$\nu$NS searches will constrain other interesting models of light new physics. For example, models of 
new nearly massless sterile neutrino-like singlet states interacting with nuclei via a baryonic current easily allow for an interaction strength
exceeding the weak interactions \cite{Pospelov:2011ha}. The oscillation of active neutrinos into 
these `baryonic neutrinos' with a long baseline will lead to a solar neutrino 
scattering signal in many DM experiments \cite{Pospelov:2012gm,Pospelov:2013rha}. 
Such signatures are very similar to the scattering of WIMP DM with a mass of a few GeV off nuclei, and the two signals can 
easily be confused. 
Constraining this class of models using CE$\nu$NS searches appears quite 
feasible \cite{Pospelov:2011ha}, and we will investigate 
such signatures in detail.

The rest of this paper is organized as follows. In the next Section we define the two classes of models to be studied. 
In Section~3, we present some calculational details of the production and detection of light states in fixed target experiments. 
Then in Section~4, we discuss the sensitivity reach of the planned CE$\nu$NS experiments at the SNS and the BNB to these new light states,
and present our conclusions in Section~5.

\section{Light dark states}
\label{sec:ldm}

Light thermal relic dark matter, with mass below a few GeV, generically requires new annihilation channels with light mediators in order not to over-close the universe. The simplest mediators couple via the renormalizable portal interactions. We will study one benchmark model below, which uses the vector portal. We also consider new massless SM-singlet sterile-neutrino-like states which couple to a gauged baryonic current.

\subsection{Benchmark model for light dark matter }

The model we study uses a spontaneously broken U(1)$'$ gauge symmetry in the hidden sector, leading to a massive vector $V_\mu$ which is kinetically mixed with the photon~\cite{Holdom:1985ag}, and dark matter is a hidden scalar or fermion $\ch$ charged under $U(1)'$. At low energies, the Lagrangian is given by
\begin{align}
\label{eq:L1}
{\cal L} &=  {\cal L}_\chi - \frac{1}{4}V_{\mu\nu}V^{\mu\nu} + \frac{1}{2}m_V^2 V_\mu V^\mu - \frac{\kappa}{2} V^{\mu\nu} F_{\mu\nu} + \cdots \;\;\;\;\; 
\end{align}
with
\begin{align}
{\cal L}_\chi & = 
\begin{cases}
i \bar \chi \not \!\! D \chi - m_\chi \bar \chi \chi,  ~~~~~~~ ({\rm Dirac ~ fermion ~ DM})\\
|D_\mu \chi|^2 - m^2_\chi |\chi|^2,~~~~({\rm Complex ~ scalar ~ DM})
\end{cases} \nonumber 
\end{align}
where $D = \partial - i g' q_e V$, with $g'$ ($q_e$) the $U(1)'$ gauge coupling (charge), and the ellipsis denotes terms associated with the spontaneous breaking of $U(1)'$, which will not be important here. 

While we will consider kinematics with both on- and off-shell mediators, it is important to note that when the vector $V$ can decay on-shell to dark matter, $m_V > 2m_\ch$, a light complex scalar DM candidate is less constrained by the impact of annihilation on the CMB, as it is $p$-wave suppressed. Other ways to evade the CMB constraint with fermionic DM include a
particle-antiparticle asymmetry and/or split states in the DM sector \cite{Izaguirre:2014dua}. For simplicity, we shall concentrate on the 
bosonic DM case and determine the sensitivity of future CE$\nu$NS experiment to a four-dimensional parameter space $\{g',m_V,\kappa,m_\chi\}$.
We also comment that the $m_V\to 0$ limit leads to a model of `millicharged' particles, and planned CE$\nu$NS experiments may also provide 
additional constraints in this case. However, the production of $\chi$ would then have to occur via an off-shell $U(1)'$ mediator.

\subsection{Baryonic neutrino model }

A well-motivated portal to dark states that would have a distinctly different phenomenology is the baryonic current portal. 
The gauged $U(1)_B$ baryon number current $J^\mu_B \equiv \frac{1}{3} \sum_i \bar{q}_i \gamma^\mu q_i$ is anomalous, but the anomaly can be canceled by new states 
at the electroweak scale. Therefore it can be viewed as a self-consistent low-energy limit of a larger theory. The model we  
consider involves a baryonic vector particle coupled to nucleons (through the underlying coupling to quarks) in the following way,
\be
\label{Lb}
{\cal L}_B = {\cal L}_{\chi}- \frac{1}{4}V^B_{\mu\nu}V^B_{\mu\nu} + \frac{1}{2}m_B^2 V^B_\mu V^B_\mu +
\sum_{N=n,p}i\bar N  \not \!\! D N ,
\ee
with ${\cal L}_\chi$ given by the same Lagrangian as before, with the covariant derivative 
for $U(1)_B$ having the corresponding coupling $g_B$ and charge $q_B$. We choose $q_B=1/3$ for all quarks, 
which gives $q_B=1$ for both neutron and 
proton. To make a connection with previous work, 
we note that the fermionic hidden sector field $\chi$ was denoted $\nu_b$ in \cite{Pospelov:2011ha}.
We will investigate the sensitivity of CE$\nu$NS experiments to this model as well, and determine 
the potential tests of an enhancement of the `baryonic force' relative to the 
weak force, $N_{\rm en} \equiv (g_B^2/m_{B}^2) / G_F$.

\section{Production and detection of light states}

\subsection{Fixed target production modes}

Having in mind the relatively low beam energy at the SNS, we will account for a number of production modes from meson decay. In the formulae below, we will denote the mediator collectively as $V$ for both models.

\bigskip
\noindent {\it (i)  $\pi^0$ decay in flight}

\noindent A dominant production mode in the forward direction utilizes radiative $\pi^0$ decay,
\be
 \pi^0 \longrightarrow \gamma + V^{(*)} \longrightarrow \gamma + \chi^\dagger + \chi.
\ee
If kinematically allowed, the on-shell production of $V$ is expected to dominate.
We also allow for off-shell $V^*\rightarrow \chi^\dagger \chi$ decays, which are significant when $\alpha'=(g')^2/4\pi$ (or $\al_B$) is not too small (as recently emphasized in \cite{Kahn:2014sra}), 
\be \label{eq:offshell}
  \Gamma_{\pi^0\rightarrow \gamma \ch^\dagger \ch} = \frac{1}{4\pi m_\pi} \int d\Pi_{\pi^0\rightarrow \gamma V} d\Pi_{V\rightarrow \ch^\dagger \ch} dq^2  |{\cal M}|^2.
  \ee
Here $d\Pi$ is the 2-body phase space, and \cite{Kahn:2014sra}
  \be
  |{\cal M}|^2 =  \frac{c_V\al f(q^2,p\cdot k_1,p\cdot k_2)}{\pi f_\pi^2 [(q^2-m_V^2)^2+m_V^2\Gamma_V^2]},
\ee
where the coupling takes the form,
\be
c_V =\left\{ \begin{array}{ll}  \ka^2\al\al' & {\rm for\, U(1)}' \\
               q_B^2\al_B^2   & {\rm for\, U(1)}_B  \end{array} \right.
\ee
with $f{=}(q^2{-}4m_\ch^2)(m_{\pi}^2{-}q^2)^2{-}4q^2(p\cdot k_1{-}p\cdot k_1)^2$. In these expressions $p$ is the photon momentum, $q$ the momentum of $V$, and $k_{1,2}$ the momenta of the dark sector particles in the final state, so that $q=k_1+k_2$.  

We also consider dark matter production through the decay of the $\eta$ at the BNB, which we assume follows the same momentum distribution as the $\pi^0$, but with a lower production rate $N_{\pi^0} \approx 30 N_{\eta}$ \cite{Teis:1996kx}.

\bigskip
\noindent {\it (ii) $\pi^-$ capture}

\noindent In addition to radiative pion decays, an isotropic production mode involves $\pi^-$ capture on protons, through a version of the Panofsky process
\be
 \pi^- + p \longrightarrow n + V^{(*)} \longrightarrow n + \chi^\dagger + \chi.
\ee
This mode provides an approximately monochromatic source of $V$, which gives an isotropic source of dark matter with a
`rectangular' energy distribution in the lab frame. It is particularly relevant for detectors located at large angles relative to the beam axis, as is the case for the CE$\nu$NS proposals. We utilize data from \cite{MacDonald:1976ky}, which shows a sharp radiative peak with energy $E\sim 130$~MeV.

A Monte Carlo code is used to calculate the angular acceptance and energy distribution of the dark matter particles generated through the previously discussed production channels. We sample the $\pi^0$ production distribution in order to generate $\pi^0$ decays in flight. As the $\pi^0$ distribution is not well studied for fixed target neutrino experiment energies and targets, we approximate it using charged pion parameterizations appropriate for the energies of the {\tt CENNS} and {\tt COHERENT} experiments (see \cite{amaldi1979, jaeger1974}). We utilize the Burman-Smith $\pi^+$ production distribution \cite{Burman:1989ds} for {\tt COHERENT} and the mean of the Sanford-Wang $\pi^+$ and $\pi^-$ parameterizations \cite{AguilarArevalo:2008yp} developed by MiniBooNE for {\tt CENNS}. The number of charged pions produced at the SNS is: $N_{\pi^-}=0.05\times {\rm POT}$, $N_{\pi^+} =0.17\times {\rm POT}$ \cite{Collar:2014lya}, where the number of POT for a year's running is expected to be 10$^{23}$. For the BNB, we use MiniBooNE multiplicity calculations: $N_{\pi^-} \approx N_{\pi^+} = 0.9\times {\rm POT}$ \cite{AguilarArevalo:2008yp}, and assume $10^{21}$ POT in total.

\subsection{Nuclear Scattering Rate}

Depending on the momentum exchanged, scattering off nuclei can be either coherent or incoherent. Given that coherent scattering dominates for low momentum transfer, it is simplest to model the transition with a form-factor as is used in direct detection. 

For incoherent scattering off nucleons, the leading term in the cross section has the form,
\be \label{eq:scatter}
 \frac{d\si_{\ch N}}{dE_{\ch}} = 4\pi c_V F_1(Q^2) \frac{2m_N E E_\ch{-}m_\ch^2 (E{-}E_\ch)}{(E^2-m_\ch^2)(m_V^2+Q^2)}{+}\cdots
 \ee
where $E_\ch$ is now the energy of the recoiling DM particle, while $Q^2=2m_N(E-E_\ch)$ is the momentum transfer. $F_1$ is a charge form factor,
\be
F_1(Q^2) = \left\{ \begin{array}{ll} q_{e}^{(N)} G_D(Q^2) & {\rm for\, U(1)}' \\
               G_D(Q^2)   & {\rm for\, U(1)}_B\end{array} \right.
\ee
with $G_D(Q^2)$ the Sachs form-factor, and $q_e^{(N)}$ the electric charge of the nucleon $N=p,n$. The ellipsis denotes terms associated with the nucleon dipole form-factors, which are generally subleading (for protons) and are neglected here to simplify the presentation. However, the full results incorporating these extra terms are given in \cite{deNiverville:2011it,Batell:2014yra}, and are included in the numerical analysis below.

Of particular interest in the present analysis is the possibility of coherent scattering off nuclei in the detector, for light vector masses. We can write the overall nuclear scattering cross section in the form
\be
 \si_{\rm \ch A}(E) = \int_{E_\ch^{\rm min}}^{E_\ch^{\rm max}} dE_\ch \left( f_p(q^2) \frac{d\si_{\ch p}}{d E_\ch} + f_n(q^2) \frac{d\si_{\ch n}}{d E_\ch} \right)
 \label{eq:Ascatter}
\ee
where $q=\sqrt{2M(E-E_\ch)}$ is the nuclear recoil momentum, with $E$ the initial kinetic energy. We utilize the Helm form-factor $F_{\rm Helm}(q^2)$, so that 
\begin{align}\label{eq:scatter2}
 f_{N=p,n}(q^2) &=\left\{ \begin{array}{ll}  n_N & q^2 > (50\,{\rm MeV})^2, \\
                       n^2_N |F_{\rm Helm}(q^2)| & q^2 < (50\,{\rm MeV})^2, \end{array} \right.
\end{align}
where $n_p=Z$, $n_n=(A-Z)$ for U(1)$'$, while $n_p=n_n=A$ for U(1)$_B$, in terms of the atomic number and mass of the nucleus.

\section{Sensitivity at SNS and Fermilab}
\label{sec:sens}

\begin{figure*}
 \centerline{
 \includegraphics[width=0.45\textwidth]{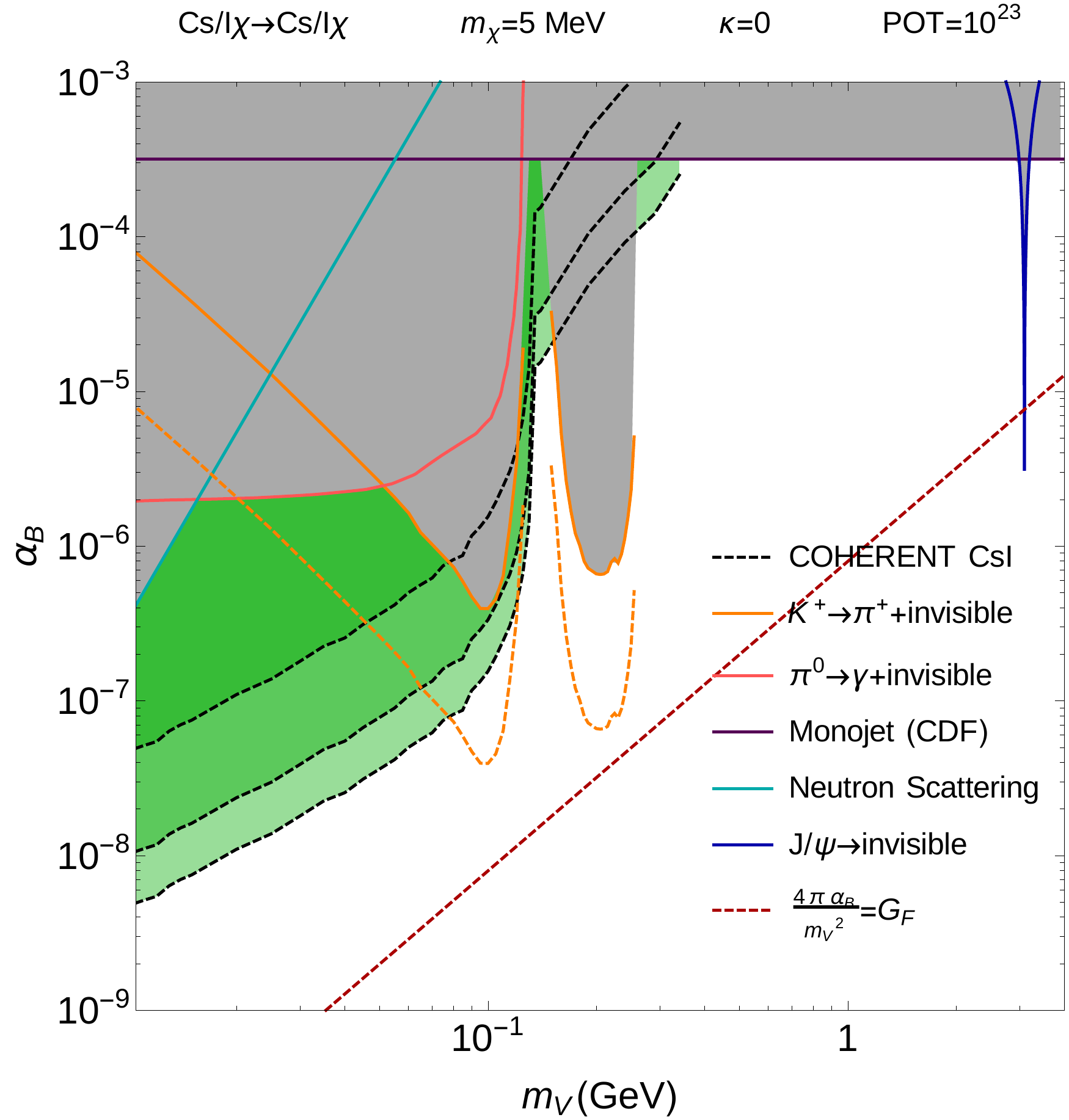}
  \hspace*{0.3cm} \includegraphics[width=0.45\textwidth]{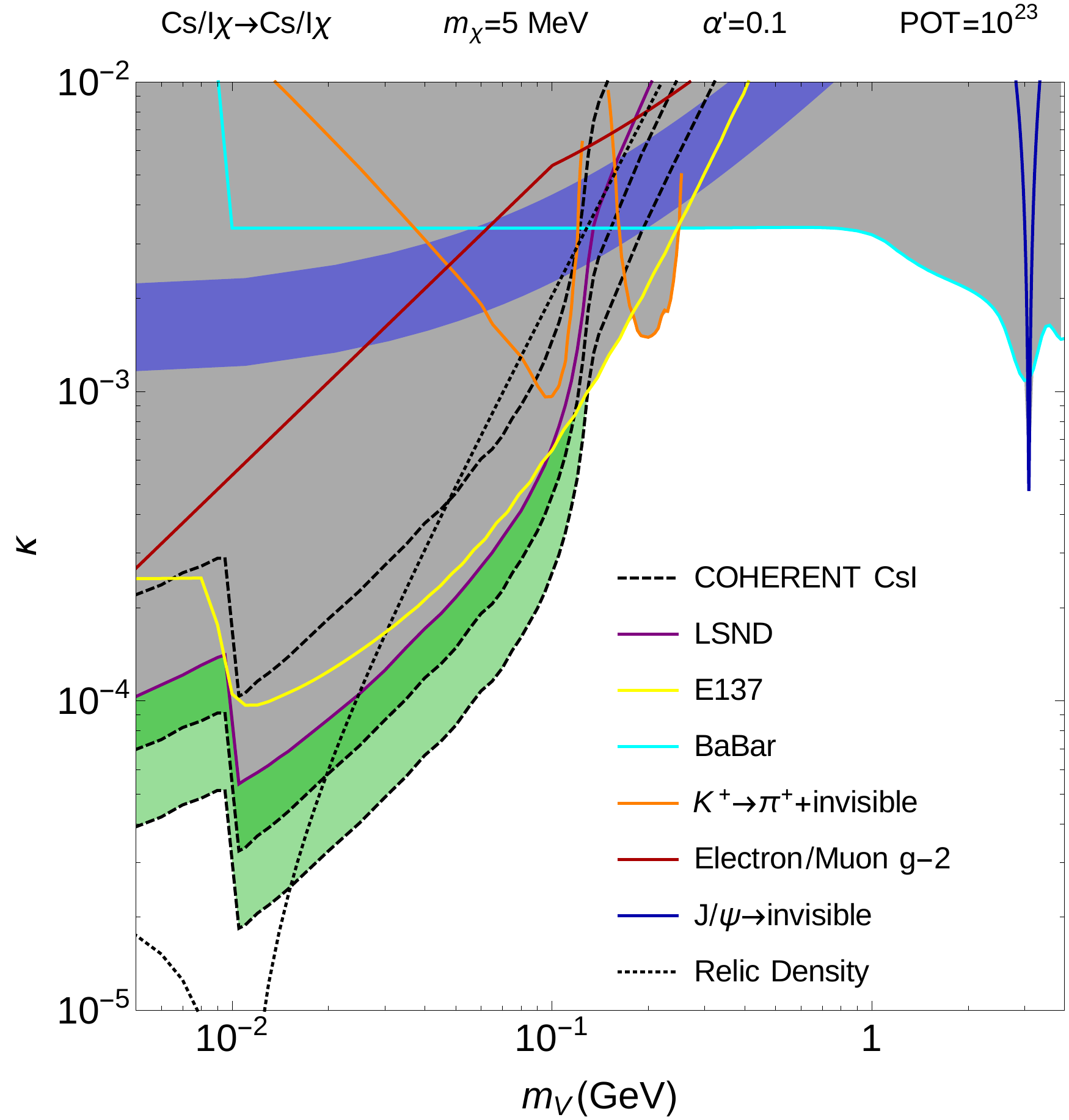}
 }
 \caption{\footnotesize Sensitivity contours for scattering events at {\tt COHERENT} (SNS), with the three green-shaded contour regions corresponding to 1 event (light), 10 events (medium) and 1000 events (dark). In grey are exclusions from other sources (see the text for further details). The left panel displays the sensitivity to the $U(1)_B$ model in the $m_V - \alpha_B$ plane, and the right panel displays sensitivity to pure vector portal DM in the $m_V-\kappa$ plane.}
 \label{fig1}
\end{figure*}

\begin{figure*}
 \centerline{
 \includegraphics[width=0.45\textwidth]{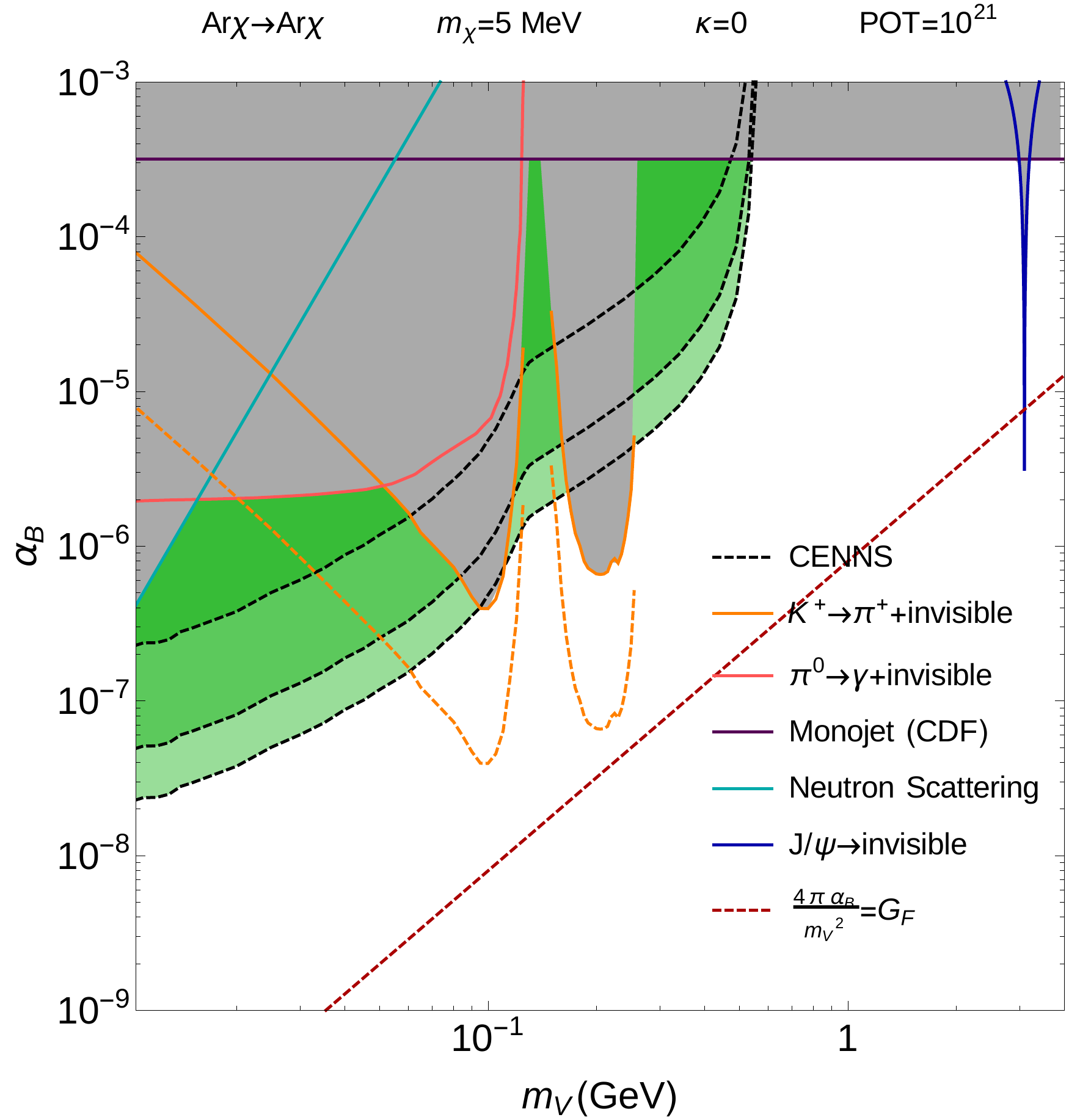}
  \hspace*{0.3cm} \includegraphics[width=0.45\textwidth]{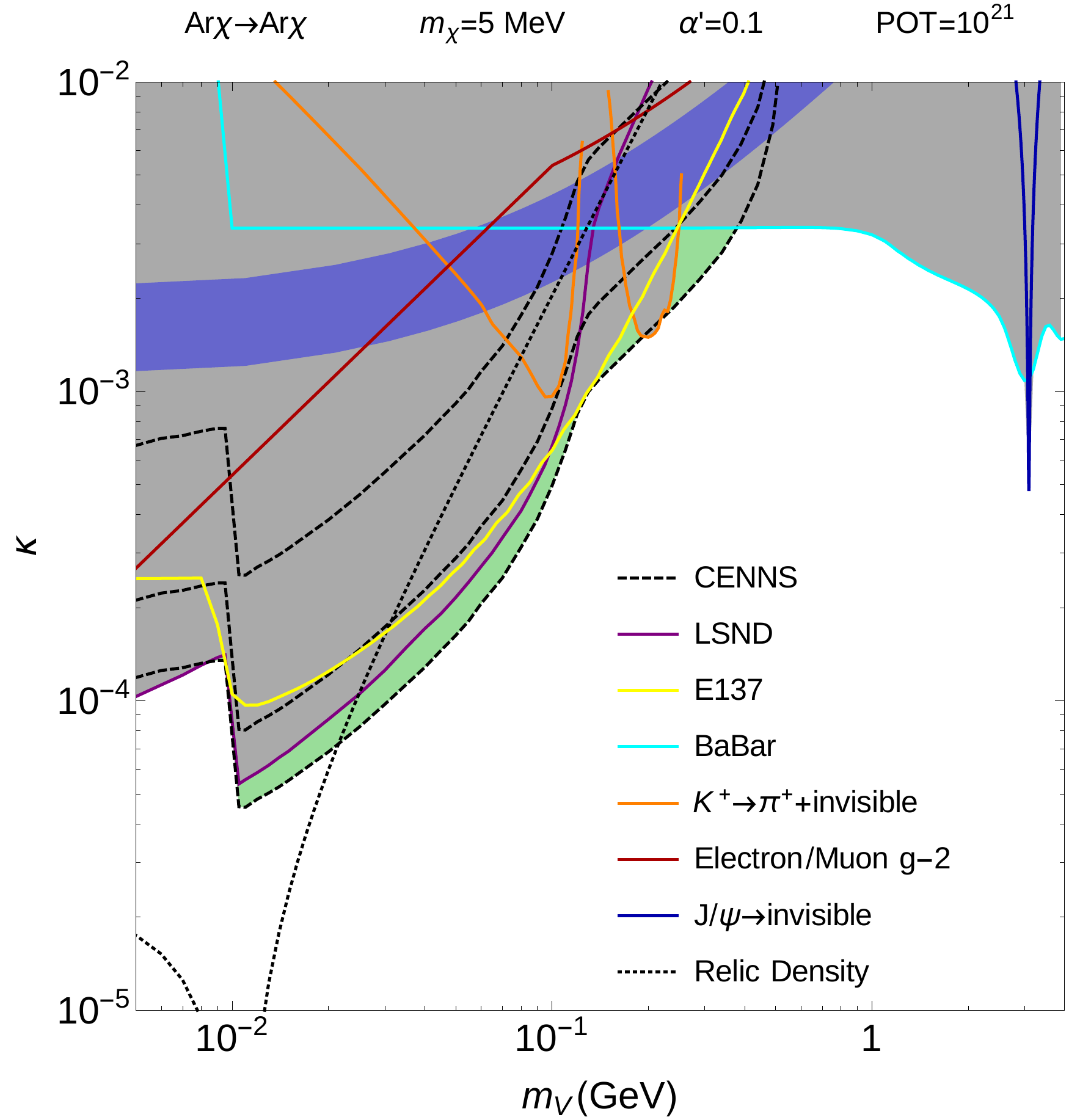}
 }
 \caption{\footnotesize Sensitivity contours are shown as for Fig.~\ref{fig1}, but for scattering events at {\tt CENNS} (Fermilab).}
 \label{fig2}
\end{figure*}

We consider similar hypothetical detector geometries for both {\tt CENNS} and {\tt COHERENT}: A cylindrical tonne-scale detector located twenty meters away from the target at a right angle to the beam-line direction. The {\tt CENNS} detector considered contains 1 tonne of ${}^{39}$Ar, while the hypothetical {\tt COHERENT} detector contains 1 tonne of CsI. The expected scattering signal was calculated by iterating over all dark matter 4-momenta generated by the production Monte Carlo that intersect the detector. Energies for recoiling nuclei are selected by sampling \eqref{eq:Ascatter}, and those that do not pass the kinematic cuts chosen for each experiment are discarded. Minimum recoil energy cuts were chosen such that the prompt neutrino background is minimized. For {\tt COHERENT}, a cut of $E_{\rm recoil}>16$ keV was adopted, and for {\tt CENNS} the cut was set at $E_{\rm recoil}>50$ keV. Incoherent nucleon scattering was also considered for $q^2 > (50\,{\rm MeV})^2$, and its contribution to the total signal was negligible for $m_V < m_{\pi^0}$. The total number of signal events can be written as
\begin{align}
\nonumber   N_{A\chi \to A\chi} &= n_A \epsilon_{\rm eff} \\
  & \times \sum_{\substack{{\rm prod.}\\{\rm chans.}}}\left(\frac{N_\chi}{2N_{\rm{trials}}}\sum_i L_i \sigma_{A\chi,i} \right),
\end{align}
where $A=$Li, Cs, Ar, and $\epsilon_{\rm eff}$ is the detection efficiency for events within the detector volume (we assume 50\%). The inner sum is over all dark matter 4-momenta $p_i$ generated by the production Monte Carlo, $L_i$ is the length of the intersection between the dark matter trajectory (with momentum $p_i$) and the detector ($L_i=0$ if the trajectory does not pass through the detector), $\sigma_{A\chi}$ is the scattering cross section \eqref{eq:Ascatter} between $\chi$ and the nucleus $A$, $N_{\rm trials}$ is the total number of decays generated by the production Monte Carlo and $N_\chi$ is the number of dark matter particles produced,
\begin{align}
 N_\chi =2 N_{\pi^0,\pi^-,\eta} {\rm Br}(\pi^0,\pi^-,\eta \to \chi \chi^\dagger X),
\end{align}
where $X$ includes any non-hidden sector end products. For the {\tt COHERENT} sensitivity curves, the signals from Lithium and Cesium nuclei are summed together.

We present the results for experiments such as {\tt COHERENT} and {\tt CENNS} in Figs.~\eqref{fig1} and \eqref{fig2} respectively. We show sensitivity contours corresponding to 1, 10 and 100 events. The computation uses cuts on the recoil energy spectrum designed to remove the coherent neutrino scattering background, and thus we anticipate that the actual sensitivity should be quite good, potentially at the ${\cal O}({\rm few})$ event level. However, a full analysis would be required to determine the full background in more detail. 

The plots also show a number of other contours that we briefly summarize below (further details can be found in \cite{deNiverville:2012ij,Batell:2014yra}). The existing constraints vary for the two model classes studied in the paper. For the vector portal dark matter model, we show fixed target constraints from LSND \cite{deNiverville:2011it} and E137 \cite{Batell:2014mga}, and missing energy constraints from BaBar \cite{aubert2008,Izaguirre:2013uxa,Essig:2013vha}. We also show constraints on the rare decay $K^+\rightarrow \pi^+ \nu\bar{\nu}$ \cite{Kplus} and on the invisible decay of $J/\Psi$ \cite{ablikim2007}. Finally, we have the constraint on $g-2$ of the electron, and the band for corrections to $g-2$ of the muon \cite{pospelov2008,bouchendira2010,hanneke2010,aoyama2012,endo2012}. In addition to the constraint contours, we exhibit a contour showing the parameters required to ensure the measured relic cosmological abundance of DM from freeze-out. This contour moves down as $\al'$ increases, as the annihilation rate is proportional to $\ka^2\al'$.

For the `baryonic neutrino' model, we also show constraints from new long-range contributions to neutron scattering, from searches for $\pi^0\rightarrow \gamma+$invisible \cite{atiya92}, and mono jet constraints from CDF \cite{shoemaker2011}. In addition to the constraint contours, we also show for reference the contour for which the baryonic interaction is of comparable strength to weak exchanges, with $4\pi \al_B/m_V^2 = G_F$. This serves to indicate the (surprising) fact that sizeable interactions via this portal are still allowed by present constraints.

\section{Concluding Remarks}
\label{sec:outlook}

We have analyzed the sensitivity of proposed coherent neutrino-nucleus scattering experiments to light dark matter and related new physics scenarios. The possibility of coherent enhancement of the new physics scattering signature allows for impressive sensitivity, particularly at the SNS, exploiting the high intensity of the proton beam. Access to this range of parameter space for light hidden sector new physics would otherwise require new dedicated experiments (see e.g. 
\cite{Izaguirre:2013uxa,Kahn:2014sra,Izaguirre:2014bca}). This provides an added motivation to pursue CE$\nu$NS experiments of this kind, beyond those associated with detecting and precisely measuring the coherent neutrino-nucleus scattering cross section. 

The sensitivity for light DM and correspondingly light ${\cal O}(10)\,{\rm MeV}$ mass vector mediators in the benchmark DM model can reach down toward kinetic mixing values as low as ${\cal O}(10^{-5})$ at SNS. This can approach the regime where astrophysical sources may provide complementary information. For example, for light vector mediators, production in the core of supernovae, via e.g. $NN\rightarrow NNV \rightarrow NN\ch\ch^\dagger$, can provide constraints from excess cooling that set in for sufficiently small kinetic mixing parameters \cite{Dreiner}. However, if the effective interaction strength of $\chi$ with electrons and nucleons is larger than $G_F$, as is the case here, the corresponding dark matter particles will be trapped, and unable to provide an efficient cooling channel, effectively weakening the constraints. 
It is interesting to note that another potential source of sensitivity, namely dark matter capture and annihilation (e.g. to neutrinos) in the Sun, is not competitive in the sub-GeV mass range. This is because evaporation processes limit the capture efficiency, so that the equilibrium between capture and annihilation that applies for weak-scale DM is no longer achieved.

\section*{Acknowledgements}

We would like to thank K. Scholberg for helpful discussions several years ago about CE$\nu$NS experiments. 
The work of  P.dN., M.P. and A.R. is supported in part by NSERC, Canada, and research at the Perimeter Institute 
is supported in part by the Government of Canada through NSERC and by the Province of Ontario through MEDT.

\bibliography{cscat}
\end{document}